\documentstyle[12pt]{article}
\newcommand{\eps}{\varepsilon}
\newcommand{\dsp}{\displaystyle}
\newcommand{\E}{{\cal E}}
\newcommand{\eff}[1]{_{\rm eff\,#1}}
\newcommand{\p}{{\bf p}}
\newcommand{\tth}{\tilde\theta}
\newcommand{\ra}{\rangle}
\newcommand{\tA}{{\cal A}}
\newcommand{\tF}{{\cal F}}

\begin{document}
\begin{titlepage}
\begin{center}
\large\bf Fermion in the Nonabelian Gauge Field Theory in 2+1 Dimensions
\end{center}
\vspace{0.5cm}
\renewcommand{\thefootnote}{\fnsymbol{footnote}}
\begin{center}
V.~Ch.~Zhukovsky, N.~A.~Peskov\footnote{E-mail addresses:
{\tt th180@phys.msu.su} and {\tt peskov\_\,nick@mail.ru}.}\\
{\sl Physical Faculty, Moscow State University,\\
119899, Moscow, Russia}
\end{center}
\begin{abstract}
The massive $SU(2)$ gauge field theory coupled with fermions is considered
in
$2+1$ dimensions. Quark energy spectrum and radiative shift in constant
external nonabelian field, being exact solution of the gauge field
equations
with the Chern-Simons term, are calculated. Under the condition $m =
\theta/4$ the quark state is shown to be supersymmetric.
\end{abstract}
\vspace{0.3cm}
\vfill
\end{titlepage}

Gauge theory models in $(2+1)$-space-time are useful in developing ideas
for
four-dimensional models, such as their high temprature behavior, boundaries
of
the spatial region occupied by gauge field etc. In this paper, we discuss
topologically massive $SU(2)$ gauge theory coupled with fermions and
compute
one-loop correction to quark energy.

The total Lagrangian of three-dimensional topologically massive
$SU(2)$-gluodynamics of potentials $\tA_{\mu}\equiv\tau^a\tA_{\mu}^a/2$,
where
$\tau^a$ are Pauli matrices in color space, and quarks is described as
follows~\cite{CSterm}
\begin{equation}
{\cal L} = -\frac14\tF_{\mu\nu}^a\tF^{a\,\mu\nu}-\frac\theta4
\eps^{\mu\nu\alpha}\left(\tF_{\mu\nu}^a\tA_{\alpha}^a-\frac{g}3
\eps^{abc}\tA_{\mu}^a\tA_{\nu}^b\tA_{\alpha}^c\right)-
\frac1{2\xi}\left(\nabla_\mu^{ab}a^{b\,\mu}\right)^2+
{\overline\psi}(D_\mu\gamma^\mu-m)\psi,
\label{nLagrQCD}
\end{equation}
where $\mu,\nu = 0,1,2$; $a = 1,2,3$, the color field tensor is given by
$\tF_{\mu\nu}^a = \partial_\mu^{}\tA_{\nu}^a-\partial_\nu^{}\tA_{\mu}^a+
g\eps^{abc}\tA_{\mu}^b\tA_{\nu}^c$. The total gauge field $\tA^{a\,\mu}$ is
represented as the sum of the classical background field and quantum
fluctuations $a^{a\,\mu}$, i.e.
$$
\tA^{a\,\mu} = A^{a\,\mu}+a^{a\,\mu},
$$
so that $\nabla_\mu^{ab} = \delta^{ab}\partial_\mu+g\eps^{abc}A^c_\mu$ is
the
background covariant derivative and $D_\mu^{ab} = \delta^{ab}\partial_\mu+
g\eps^{abc}\tA^c_\mu$ is the total covariant derivative. The third term in
(\ref{nLagrQCD}) is the gauge-fixing term. The coefficient $\theta$ in
front
of the second term in (\ref{nLagrQCD}) (Chern-Simons term) is the
Chern-Simons
(CS) mass of the gauge field. We use the following representation for the
$\gamma$-matrices in $2+1$ dimensions: $\gamma^0 = \sigma^3$, $\gamma^{1,2}
=
i\sigma^{1,2}$, with $\sigma^i$ as Pauli matrices. The $\gamma$-matrices
obey
the following relation: $\gamma^\mu\gamma^\nu = g^{\mu\nu}-
i\eps^{\mu\nu\alpha}\gamma_\alpha$.

It was shown in \cite{hep} that  the classical constant nonabelian
potentials
\begin{equation}
A^{a\,\mu} = \frac{\theta}{2g}\delta^{a\mu}\chi_{\lambda\omega}^{(a)},
\label{n-field}
\end{equation}
with normalized constant vector $\chi_{\lambda\omega}^{(a)} = (\lambda i,\,
\lambda\omega i,\,\omega)$ satisfy the field equations with the
Chern-Simons
term without external currents. In (\ref{n-field}) $\lambda = \pm1$ and
$\omega = \pm1$ take its values independently. The Kronecker delta
$\delta^{a\mu}$ in (\ref{n-field}) implies that directions 1, 2, 3 in the
color space correspond to directions 1, 2, 0 in the Minkowski $2+1$
space-time, respectively. In what follows, like in ~\cite{hep}, where
corrections to the gluon energy were considered, these solutions are chosen
as
the background. 

Considering the one-loop corrections, it is sufficient to retain only the
terms in the Lagrangian (\ref{nLagrQCD}) quadratic in the quantum fields.
They
determine the quark energy spectrum in the gauge field (\ref{n-field})
$$
\eps_1^2=\p^2+m\eff1^2,\qquad\eps_2^2=\p^2+m\eff2^2,
$$
where
\begin{equation}
m\eff1^2 = (m-\tth)^2,\qquad m\eff2^2 = (m-\tth)(m+3\tth)
\label{effect}
\end{equation}
and $\tth =\theta/4$. These branches of the energy spectrum correspond to
the
plane-wave solutions $\psi_s(x) = \exp(-i\eps_s t+i\vec{px})\Psi_s,$ with
$s =
1,2,$ and $\Psi_s$ as constant spinors, of the Dirac equation
$$
[\gamma^\mu(p_\mu+gA_\mu)-m]\psi = 0,
$$
and are related to two opposite projections of the particle color spin
operator~\cite{hep}
\begin{equation}
{\bf J} = J^a\tau^a/2 =
\tth^{-1}gA^\mu p_\mu- \omega\gamma^0\tau^3(p_\mu\gamma^\mu+\tth-m),
\label{nJ}
\end{equation}
defined as eigenvalues of the equation
$$
{\bf J}\Psi_{s} =
(-1)^s(\tth-m)\Psi_{s}.
$$
We note, that the r.h.s.\ of the last equation vanishes, when $m = \tth$.
In
this special case, the quark effective masses in this gauge field
(\ref{effect}) are also equal to zero, and the two branches of the energy
spectrum coincide. For values of the mass lying in the interval
$-\tth-2|\tth|<m<-\tth+2|\tth|$, the energy squared, $\eps_2^2$, becomes
negative for certain values of the quark momentum $\p$, and tachyonic modes
arise in the quark spectrum. Moreover, it appears that $\Psi_1 = \Psi_2$
under
this condition. 

We now consider the one-loop radiative shift of the quark energy. According
to~\cite{radshift}, the quark energy radiative correction $\Delta\eps$ is
obtained by averaging the mass operator over the quark state in the
external
field $A^\mu$, specified by equation (\ref{n-field}). The one-loop
radiative
correction to the quark energy is given by formula
$$
\Delta\eps = -\frac{i}{T}\int\!\!\int d^3\!xd^3\!x'\,
\bar{\psi}_k(x)M_{kl}(x,x')\psi_l(x'),
$$
where
$$
iM_{kl}(x,y) = -ig^2\gamma^\mu(\tau^a/2)_{kn}S_{nm}(x,y)\gamma^\nu
(\tau^b/2)_{ml}D^{ab}_{\mu\nu}(x,y)
$$
is the one-loop mass operator. The expression for $\Delta\eps$ has to be
applied with due regard to the fact that the quark Hamiltonian in the
background (\ref{n-field}) becomes non-Hermitian. In the momentum
representation we have 
$$
\begin{array}{c} \Delta\eps =
\bar{\Psi}_k(p)iM_{kl}(p)\Psi_l(p),\\[2ex]\dsp iM_{kl}(p) =
ig^2\gamma^\mu(\tau^a/2)_{kn}\int\!d^3\!k\,S_{nm}(p-k)
\gamma^\nu(\tau^b/2)_{ml}D^{ab}_{\mu\nu}(k).
\end{array}
$$
In order to calculate the quark mass operator in the external field the
quark
and gluon propagators have to be found. The quark Green's function
satisfies
the equation 
$$
[\gamma^\mu(i\partial_\mu+gA_\mu)-m]S(x,y) = \delta(x-y).
$$
In the background gauge field $A^\mu$ considered above, the quark Green's
function in the momentum representation has the form
$$
\begin{array}{c}
S(p) = [(p^2-m\eff1^2)(p^2-m\eff2^2)]^{-1}
\left\{(p^2-(m-\tth)^2)[\gamma^\mu(p_\mu+gA_\mu)+m]-\right.\\[2ex]
\left.2(\gamma^\mu p_\mu+m-\tth)[gA^\nu p_\nu+\tth(m-\tth)]\right\},
\end{array}
$$
with the quark effective masses $m\eff1$ and $m\eff2$ defined above.

The gluon propagator in the external field in the gauge $\xi = 1$ takes the
form
$$
\begin{array}{c}\dsp D^{ab}_{\mu\nu} =
\delta^{ab}g_{\mu\nu}\left[\frac1{2\E}+\frac\E{2\alpha}\right]-
\frac{4g^2}{\theta^2}A^a_\mu
A^b_\nu\left[\frac1{3\E}-\frac\E{3\beta}\right]+
\frac{4g^2}{\theta^2}A^a_\nu
A^b_\mu\left[\frac1{2\E}-\frac\E{2\alpha}\right]+
\\[2.2ex]\dsp
\frac{16g^2\E}{\theta^2\alpha\beta}F^a_{\mu\alpha}F^b_{\nu\beta}p^\alpha 
p^\beta+
\frac{8ig^2}{\theta^2\beta}(F^a_{\mu\alpha}A^b_\nu-F^b_{\nu\alpha}A^a_\mu)
p^\alpha+
\frac{16ig^2}{\theta^3\alpha}\eps^{\alpha\beta\gamma}F^a_{\mu\alpha}
F^b_{\nu\beta}p_\gamma,
\end{array}
$$
where we used the notations $\E = p^2+\frac12\theta^2$, $\alpha = \E^2-
4\theta^2p^2$ and $\beta = \E^2-6\theta^2p^2$. 

In the case of an arbitrary quark mass value $m\neq\tth$, the quark has two
different color states. It is interesting to consider the special case,
when
$m=\tth$, and, as it has been mentioned above, the quark effective masses
vanish. We remind that under this condition, $m\eff1 = m\eff2 = 0$, only
one
quark state survives. The corresponding plane wave solution of the Dirac
equation $\psi(x) = (2\pi)^{-1}\exp(-i\eps t+i\vec{px})\Psi(p)$ is
determined
by the constant spinor
$$
\Psi(p) =
\frac18\left(\begin{array}{c}
[\lambda(\kappa-1)+(\kappa+1)][(\omega+1)e^{i\phi}+(\omega-1)]\\[1.5ex]
-[\lambda(\kappa-1)-(\kappa+1)][(\omega-1)e^{-i\phi}+(\omega+1)]\\[1.5ex]
[\lambda(\kappa+1)-(\kappa-1)][(\omega-1)e^{i\phi}+(\omega+1)]\\[1.5ex]
[\lambda(\kappa+1)+(\kappa-1)][(\omega+1)e^{-i\phi}+(\omega-1)]
\end{array}\right).
$$
It should be emphasized that in this formula the quark momentum $\p$ is
assumed to be nonzero. Here $\kappa = \pm1$ is the sign of the energy
($\eps\equiv p_0 = \kappa|\p|$) and the phase $\phi$ is defined by the
relation $p^2\pm ip^1 = |\p|e^{\pm i\phi}$. We used representation for the
Dirac operator and the Hamiltonian, where $\gamma$-matrices are inserted
into
isospin color Pauli matrices $\tau^a$. It should be also emphasized that,
since the fermion Hamiltonian is non-Hermitian, it provides only {\it one}
solution of the Dirac equation for one value of the energy sign.

After integration over intermediate momentum $k$ the mass operator is
expressed in the form 
$$
iM(p) = ig^2\frac{\pi^2\sqrt2}{|\theta|}\left[
-b_1\theta+\frac59\left(\frac2\theta gA^\mu+\frac12\gamma^\mu\right)p_\mu-
gA^\mu\gamma^\nu(b_2\,g_{\mu\nu}-b_3\,\theta^{-2}p_\mu p_\nu)\right],
$$
where
$$
\begin{array}{c}\dsp
b_1 = \frac56-\frac{i}{12}(3\sqrt2+4\sqrt3),\\[2ex]\dsp
b_2 = \frac{10}9+\frac{i}{9}(8\sqrt3+9\sqrt2),\\[2ex]\dsp
b_3 = \frac{4}{45}-\frac{4\sqrt3i}{15}.
\end{array}
$$

Now we can find the one-loop radiative correction to the quark energy in
the
external field, which turns out to be vanishing in this particular state:
$$
\Delta\eps =
\frac{5\pi^2\sqrt2}{18}ig^2|\theta|^{-1}
\left(p_0-\kappa\sqrt{(p^1)^2+(p^2)^2}\right)\equiv0.
$$
It should be emphasized that this result is valid only for finite values of
the quark momenta $\p = (p^1,\,p^2)$. For the case of vanishing $\p$, a
special consideration is needed.

The fact that a fermion remains effectively massless even in the one-loop
approximation signals the presence of a certain symmetry of the problem in
question. In fact, the results obtained demonstrate the supersymmetry
property
of the state considered.

As is well known (see, e.g.,~\cite{zhuk}), the minimal representation of
SUSYQ
(SUSY Quantum Mechanics) is provided with the supercharges $Q_1$, $Q_2$ and
the Hamiltonian $H_S$:
$$
H_S = Q_1^2 = Q_2^2,\qquad[H_S,Q_i] = 0,\qquad\{Q_i,Q_j\} =
2\delta_{ij}H_S,
\quad i,j=1,2.
$$

The quark Hamiltonian in the external field (\ref{n-field}) can be written
in
the form
$$
H = ip^1\gamma^2-ip^2\gamma^1+\tth(\gamma^0-\omega\tau^3)-
\lambda\tth(\tau^1\gamma^2-\omega\tau^2\gamma^1).
$$
The SUSYQ Hamiltonian $H_S = H^2$ is found to be
$$
H_S = \p^2+2\tth[i\lambda(p^1\tau^1+\omega p^2\tau^2)-
i\omega\tau^3(p^1\gamma^2-p^2\gamma^1)].
$$
It is interesting to note that $H_S$ can be written in the form
$$
H_S = \p^2-2\tth{\bf J},
$$
where $\bf J$ is the operator defined in (\ref{nJ}). Under the condition $m
=
\tth$ we have ${\bf J}^2 = 0$. The supercharges, corresponding to the above
Hamiltonian $H_S$, can be found:
$$
\begin{array}{c}\dsp
Q_1 = \left[\sqrt2\gamma^0-\frac{i\lambda\omega\tth}{|\p|}\right]
(p^1\gamma^1+p^2\gamma^2)-
\frac{\omega\tth}{|\p|}\gamma^0\tau^3(\omega
p^1\tau^1+p^2\tau^2)+\\[2.5ex]\dsp
\frac1{|\p|}(p^1\gamma^1+p^2\gamma^2)(\omega
p^1\tau^1+p^2\tau^2),\\[2.5ex]\dsp
Q_2 =
\left[\frac{i\omega\sqrt2}{|\p|}+\frac{\lambda\tth}{\p^2}\gamma^0\right]
(p^1\gamma^1+p^2\gamma^2)(\omega p^1\tau^1+p^2\tau^2)+
i\tth\tau^3+i\omega\gamma^0(p^1\gamma^1+p^2\gamma^2).
\end{array}
$$

It is interesting to consider the quark ground state, when its momentum is
equal to zero. Solution of the Dirac equation is easy to find in this case.
It
has the form 
\begin{equation}
\Psi(p=0) =
\frac{c_1}{\sqrt2}\left(\begin{array}{c}
\omega+1\\\omega-1\\0\\0\end{array}\right)+
\frac{c_2}{\sqrt2}\left(\begin{array}{c}
0\\0\\\omega-1\\\omega+1\end{array}\right)+
\frac{c_3}{\sqrt8}\left(\begin{array}{c}
\omega-1\\\omega+1\\\lambda(\omega+1)\\\lambda(\omega-1)\end{array}\right),
\label{zero}
\end{equation}
where constants $c_i$ are arbitrary up to a normalization condition.

In order to demonstrate the supersymmetry property of the Dirac equation
$D\psi = 0$, where
$$
D = \gamma^\mu(p_\mu+gA_\mu)-m
$$
is the Dirac operator, we consider its zero-mode solutions with $p^0 = 0$
under the condition $m = \theta/4$ and with vanishing fermion mass in
virtue
of the condition $gA^0\gamma^0 \psi = m\psi$, which means
$\sigma^3\tau^3\psi
= \omega\psi$. In this case the Dirac equation takes the form 
$$
-(\gamma^1\nabla^1+\gamma^2\nabla^2)\psi_0 = 0.
$$
Then we introduce combinations 
$$
b^\pm = (2i)^{-1}(\gamma^1\pm i\gamma^2),
$$
which play the role of fermionic anticommuting operators (creation and
annihilation operators). The Dirac operator has the form
$$
D = Q_++Q_-,
$$
where $Q_+ = 2b^+\nabla_u$, $Q_- = 2b^-\bar\nabla_u$ while $\nabla_u =
\partial_u-igA_u$, $\bar\nabla_u = \bar\partial_u-ig\bar A_u$, and here
$\partial_u = \frac12(\partial_1-i\partial_2)$, $\bar\partial_u =
\frac12(\partial_1+i\partial_2)$, etc. The standard SUSY hamiltonian has
the
form 
$$
H_S = (D^2) = \{ Q_+,Q_-\},\qquad Q_\pm^2 = 0,\qquad [H_S,Q_\pm] = 0.
$$

We set $\psi_0 = \psi_\tau\psi_s$, where $\psi_\tau$ describes the color
(boson) state and $\psi_s$ specifies the spin (fermion) state. For the
ground
(zero-mode) state we have: $H_S\psi = 0$. This means 
$$
Q_+\psi_0 = 0,\quad Q_-\psi_0 = 0.
$$
Consider $Q_+\psi_0 = 0$, or $b^+D_u\psi_0 = 0$. Let $\psi_s = |0\ra_s$,
then
$b^+\psi_s = |1\ra_s$ and then $D_u\psi_\tau = 0$ ($|0\ra$ and $|1\ra$ are
fermionic states). For $\partial_{1,2}\psi_0 = 0$ we have
$$
(\tau^1-i\omega\tau^2)\psi_\tau = 0.
$$
Then if $\omega = +1$, we recall that $\tau^3\sigma^3 = \omega = +1$, and
for
$\sigma^3\psi_s = -\psi_s$, we have $\tau^3\psi_\tau = -\psi_\tau$ and
$(\tau_1-i\tau_2)\psi_{\tau} = 0$. Hence $\psi_\tau = |0\ra_\tau$. Excited
solutions are constructed in a standard way. Now it is clear that the first
two terms in (\ref{zero}) describe the states with the SUSY property.

\vspace{3ex}\noindent
One of the authors (N.A.P.) gratefully acknowledges the hospitality of
Prof.
M\"uller-Preu\ss ker, Prof. Ebert and their colleagues at the particle
theory
group of the Institut f\"ur Theoretische Physik, Humboldt Universit\"at zu
Berlin extended to him during his stay there.

\end{document}